\documentclass[a4paper,11pt]{article}
\usepackage{aaskaiid}
\usepackage{graphicx}
\usepackage{soul}
\usepackage{orcidlink}
\usepackage[ragged]{sidecap}
\usepackage{xcolor}

\title{Tracing gas outflows in molecular gas-rich galaxies with the SKA}
\ShortTitle{Tracing gas outflows in molecular gas-rich galaxies}

\author[1]{Mamta Pandey-Pommier\orcidlink{https://orcid.org/0000-0001-5829-1099}}
\ShortName{Pandey-Pommier et al.} 
\author[2]{Ramya Sethuram\orcidlink{https://orcid.org/0009-0005-9999-132X}}
\author[3]{Chiranjib Konar\orcidlink{0000-0002-2530-3812}}
\author[2]{Chinnathambi Muthumariappan}
\author[4]{Subhashis Roy\orcidlink{https://orcid.org/0000-0002-7779-8909}}
\author[5]{Nirupam Roy\orcidlink{https://orcid.org/0000-0001-9829-7727}}
\author[6]{Alexandre Marcowith\orcidlink{https://orcid.org/0000-0002-3971-0910}}

\affiliation[1]{Pole Scientific, University Catholic of Lyon- University of Lyon, 10 place des Archives 69288, Lyon, France}
\emailAdd{mamtapommier@gmail.com}
\affiliation[2]{Indian Institute of Astrophysics, II Block, Koramangala, Bangalore 560034, India}
\emailAdd{ramya@iiap.res.in} 
\affiliation[3]{Department of Physics, Amity Institute of Applied Sciences, Amity University Uttar Pradesh, Sector-125, Noida 201313, U.P., India} 
\emailAdd{chiranjib.konar@gmail.com}
\affiliation[2]{Indian Institute of Astrophysics, II Block, Koramangala, Bangalore 560034, India}
\emailAdd{muthu@iiap.res.in}
\affiliation[4]{National Centre for Radio Astrophysics - Tata Institute of Fundamental Research, \\
Ganeshkhind, Pune 411007, Maharashtra, India}
\emailAdd{roy@ncra.tifr.res.in}
\affiliation[5]{Indian Institute of Science, CV Raman Road, Bangalore-560012, India}
\emailAdd{nroy@iisc.ac.in}
\affiliation[6]{CNRS/Laboratoire Univers et Particules de Montpellier, Universit\'e de Montpellier \\
LUPM CC 072 - Place Eug\`ene Bataillon 34095 Montpellier Cedex 5, France}
\emailAdd{alexandre.marcowith@umontpellier.fr}

\abstract{
Active galactic nuclei with powerful radio jets play a key role in galaxy evolution through their ability to regulate the cold gas reservoirs that fuel star formation. Jet-driven feedback can heat, compress, or expel atomic and molecular gas, thereby reshaping the interstellar medium and altering star formation efficiency. However, the physical coupling between AGN activity and the multi-phase interstellar medium remains poorly constrained, particularly in radio-loud systems where mechanical feedback and multiphase outflows are expected to dominate. SKA will provide major advances in the study of cold gas in AGN host galaxies through sensitive observations of H~{\sc i} emission and absorption, together with access to selected low-frequency molecular transitions within the SKA~1 frequency range, including OH, H$_2$CO, CH$_3$OH, and, at high redshift, low-$J$ transitions of CO, HCN, and HCO$^{+}$ in rare bright systems. Combined with radio continuum measurements, these tracers will provide direct constraints on gas mass, kinematics, turbulence, and inflow/outflow signatures, enabling detailed studies of feedback-regulated cold gas reservoirs in AGN environments.
In this chapter, we examine how SKA1 observations of neutral hydrogen, complemented by molecular-line and radio continuum studies, can be used to quantify multiphase gas flows and feedback energetics in molecular-gas-rich radio galaxies. SKA surveys will enable population-level studies of AGN-driven feedback, providing a new framework for understanding how radio jets regulate the cold interstellar medium and star formation across cosmic time.}

\begin{document}
\maketitle

\section{Introduction: AGN-driven regulation of cold gas in galaxies}
The evolution of galaxies is strongly linked to the thermal and dynamical state of their cold interstellar medium (ISM), which provides the raw material for star formation and black hole growth. In massive galaxies, this gas reservoir is continuously reshaped by both environmental processes and internal energy injection mechanisms. In dense environments such as galaxy groups and clusters, ram-pressure stripping, tidal interactions, and strangulation can remove or redistribute cold gas, altering the efficiency with which galaxies form stars \citep{Boselli2006}. In addition, radio-loud AGNs inject mechanical and radiative energy into the surrounding ISM through jets, winds, and shocks, thereby modifying the structure, kinematics, and phase balance of the gas. Despite extensive work, the coupling between radio jet-mode feedback and the multi-phase ISM remains incompletely understood.

Understanding AGN-driven feedback and regulation of star formation, therefore, requires direct probes of the cold gas phases that fuel and sustain star formation. These phases are effectively traced through observations of H {\sc i} gas and molecular species such as CO, OH, and HCO$^{+}$, as well as radio recombination lines (RRLs). Together, these tracers provide a means of investigating cold gas in both emission and absorption, including heavily obscured regions at optical and infrared wavelengths \citep{Kennicutt2012}. These diagnostics enable measurements of gas mass, kinematics, turbulence, and star formation activity, providing a unified framework for investigating feedback processes and galaxy evolution across cosmic time \citep{saintonge2022}. In particular, H~{\sc i} traces extended atomic gas reservoirs through the 21-cm line in both emission and absorption, while molecular gas is commonly traced through low-$J$ CO transitions and dense gas tracers such as HCN and HCO$^{+}$ \citep{Krumholz2013, Leroy2008}. Together, these probes provide a multi-phase view of the cold gas cycle and its role in regulating star formation. The formation and distribution of cold gas are themselves governed by local physical conditions, including metallicity, dust content, radiation field intensity, and ambient pressure. In star-forming discs, H~{\sc i} is typically distributed over extended outer regions, while molecular gas forms within denser, shielded environments through grain-surface chemistry and is subsequently traced by CO emission \citep{Krumholz2013, Leroy2008}. Gas-rich galaxies generally contain $\sim10^{9},M_\odot$ of atomic gas, with the H~{\sc i}-to-H$_2$ conversion regulated by pressure and shielding. H~{\sc i} emission has been detected up to $z\sim0.4$ in deep surveys and stacking analyses, while absorption studies extend to $z\sim3.37$ \citep{fernandez2016, Morganti2015, aditya2018}. Similarly, CO(J=1--0) emission has now been detected up to $z\sim6$ using ALMA and NOEMA \citep{saintonge2022}.

Further, these multi-phase tracers can be used to investigate how cold gas responds to both environmental processes and AGN-driven feedback in dense and dynamically active systems. In galaxy clusters and interacting systems, gas reservoirs are frequently affected by environmental quenching processes that can remove, heat, or redistribute the cold ISM. However, these same processes may also compress gas and enhance central molecular gas formation, depending on the balance between stripping, cooling, and pressure confinement \citep{Volk1996, Boselli2006}. In radio-loud FR I/II systems, AGN-driven jets introduce an additional dominant source of energy injection into the ISM. Jet–ISM interactions drive shocks, turbulence, and entrainment across atomic and molecular phases, leading to multiphase outflows and strongly disturbed gas kinematics \citep{cicone2014, Morganti2018}. These systems can nevertheless retain substantial molecular gas reservoirs ($>10^{10},M_\odot$), which often coexist with high-velocity outflows traced through CO and OH transitions \citep{Castignani2020, Oosterloo2017, Hardcastle2007}. CO emission originates in UV-shielded regions where ionised carbon reacts with OH and is detectable through its rotational transitions out to $z \sim 7$ with ALMA and the JVLA \citep{Carilli2013, Tacconi2020}. Such environments therefore provide key laboratories for studying how environmental and AGN-driven processes jointly regulate the survival, excitation, and redistribution of cold gas in radio galaxies.
\begin{figure}[ht!]
    \centering
    \includegraphics[width=0.7\textwidth]{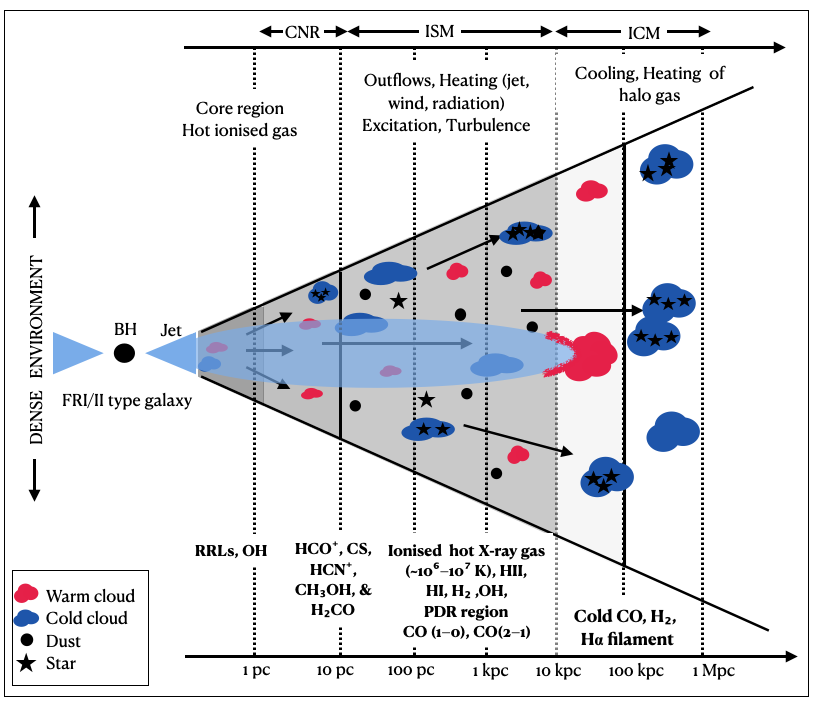} 
    \caption{Schematic representation of the multi-phase interstellar and circumgalactic medium in AGN host galaxies, illustrating the different spatial scales over which AGN feedback operates. The diagram highlights the circumnuclear region (CNR), the host galaxy interstellar medium (ISM), and the surrounding intra-cluster medium (ICM), and shows how jet-driven feedback influences the molecular, atomic, and ionized gas phases. These processes regulate gas cooling, turbulence, and the formation of complex molecules within the ISM. Adapted from \citep{Harrison2024}.}
    \label{fig:gas-phase-moheg}
\end{figure}


In addition to CO, H~{\sc i}, and H$_2$, the cold ISM contains other rich molecular species, including OH, HCN, HCO$^{+}$, and $CH_{3}OH$, which trace dense gas, shocks, and chemical processing in galaxies and their environments (see Fig.~\ref{fig:gas-phase-moheg}; \citep{Oosterloo2017}). OH megamaser emission is particularly detected through its 18-cm transitions in dusty starbursts and galaxy mergers out to redshifts of $z \sim 0.1$, as demonstrated by Arecibo surveys \citep{Darling2001, Darling2007}. Dense molecular regions ($n > 10^4 cm^{-3}$), often associated with high pressure, star-forming clouds, are probed by molecules such as HCN and $HCO^{+}$. These species have been observed in local starbursts, AGN-host galaxies, and even in quasars at $z \sim 6.4$ with instruments such as IRAM \citep{Bertoldi2003, Riechers2007}. $CH_{3}OH$ (methanol), which predominantly forms in dusty regions via grain-surface interaction, traces shocked gas in molecular outflows or jet-ISM interactions.  \citep{Impellizzeri2008} observed $CH_{3}OH$ in absorption toward the Seyfert 2 galaxy NGC3079 (z = 0.0037) with a column density of $N_{H} > 10^{13}-10^{15} cm^{-2}$ via Effelsberg 100m telescope observations. At higher redshift, \citep{Marshall2017} detected 12.2-GHz methanol absorption in the gravitationally lensed system PKS B1830–211 at, z = 0.89 using the Long Baseline Array (LBA) at milliarcsecond resolution. The excitation and abundance of these molecules are highly sensitive to ISM conditions, including gas density, thermal balance processes, and their dust content, making them powerful tracers of feedback processes, chemical evolution, and star formation efficiency across diverse environments. While multi-wavelength observations have significantly advanced our understanding of galaxy evolution, studies of low-$J$ molecular transitions that trace the bulk molecular gas reservoir remain limited at high redshift due to the restricted sensitivity and frequency coverage of existing facilities. The SKA~1 with its broad frequency coverage (50 MHz–15 GHz), $\mu$Jy sensitivity, and wide survey capability, will transform our understanding of the interplay between AGN activity and cold gas in galaxies by enabling sensitive observations of H~{\sc i} emission and absorption, as well as redshifted low-$J$ molecular transitions such as CO, HCN, and HCO$^{+}$ in rare bright systems over a wide redshift range. These capabilities will enable direct measurements of gas mass, kinematics, and inflow/outflow signatures across statistically significant samples of the rare molecular gas-rich AGN host galaxies, opening a new parameter space for investigating mechanical feedback and the regulation of star formation across cosmic time.

In this chapter, we investigate how SKA~1 observations of neutral H~{\sc i}, in combination with selected molecular-line diagnostics and radio continuum measurements, can be used to quantify multiphase gas flows and the energetics of AGN-driven feedback in galaxy hosts. We focus in particular on low-redshift ($z \lesssim 0.2$) molecular gas-rich radio galaxies hosting extended FR I/II jets, which contain substantial multi-phase gas reservoirs and often exhibit suppressed star formation \citep{Ogle2010}. These systems are well studied across multiple wavelengths and represent a transitional phase in galaxy evolution, where both star formation and black hole accretion are regulated by jet-driven turbulence, shocks, and shear within disturbed molecular gas reservoirs. These galaxies therefore provide a natural laboratory for exploring (i) the regulation of star formation by AGN feedback, (ii) jet–ISM interactions and shock-driven heating, (iii) the role of environment in shaping cold gas content in radio-loud and cluster systems, and (iv) the dissipation of turbulence and energy in multi-phase cold gas \citep{Ogle2010}. In addition, these systems may preserve signatures of episodic jet activity, offering constraints on the duty cycle of radio-mode feedback. By combining spectral-line capabilities with continuum imaging, upcoming SKA~1 surveys will provide a statistically significant sample to conduct such studies. Overall, this chapter complements the SKA science cases of radio galaxy physics presented in the \citep{Hardcastle01.2026.SKA}, and the prospects for low-$J$ molecular transitions and high-redshift cold gas studies reviewed by \citep{RanWang01.2026.SKA} in this volume. 

\section{Jet-Driven Feedback in Molecular Gas-Rich Radio AGN}
Radio-loud AGN hosting large-scale jets and substantial molecular gas reservoirs provide direct probes of jet–ISM interaction through disturbed gas kinematics and multiphase outflows. These rare systems offer a unique view of AGN feedback and cold ISM coupling. This section focuses on how such systems trace jet-driven regulation of cold gas and star formation.

\subsection{Multi-phase Molecular Gas Outflows and Mechanical Heating in AGNs}
Understanding how AGN jets heat and redistribute molecular gas is central to quantifying how feedback regulates star formation efficiency in radio-loud galaxies. 
Nearby molecular gas-rich radio AGN are unique rare systems in which jet–ISM interactions can be traced through multi-phase spectral-line observations spanning atomic, molecular, and warm gas phases. One representative example is the class of rare Molecular Hydrogen Emission Galaxies (MOHEGs), which are radio-loud AGN characterized by unusually luminous rotational lines of molecular hydrogen (H$_2$), tracing warm molecular gas at temperatures of $T \sim 10^2$–$10^3$~K \citep{Ogle2010}. MOHEGs exhibit co-spatial multi-phase gas components: cold molecular gas traced by CO, warm H$_2$ rotational emission, and atomic gas detected through H~{\sc i} 21-cm absorption \citep{veilleux2020}. The inferred H$_2$ luminosities span $L(\mathrm{H}_2) \sim 7 \times 10^{38}$ to $2 \times 10^{42}~\mathrm{erg~s^{-1}}$, corresponding to molecular gas masses up to $\sim 2 \times 10^{10}~M_\odot$. In many cases, this warm molecular component coexists with strong PAH emission, which typically traces dust-rich star-forming regions. However, the exceptionally luminous H$_2$ emission observed in MOHEGs is difficult to explain through UV heating alone and is instead interpreted as evidence for mechanical energy injection from AGN jets through shocks, turbulence, and dissipation. This makes them ideal systems for understanding how energy from radio jets propagates across different ISM phases. 

\begin{SCfigure}[1.0][h]
    \centering
    \includegraphics[width=0.6\textwidth]{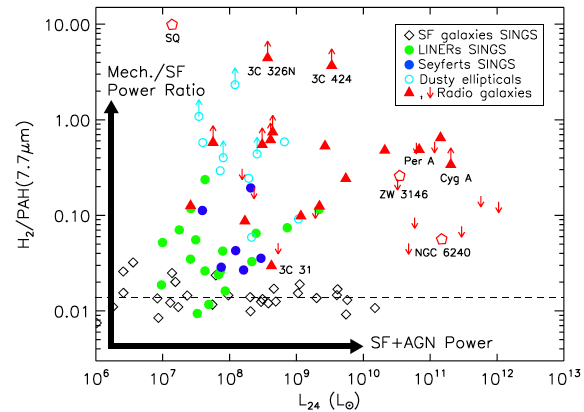}
    \caption{Ratio of H$_2$ (S(0)–S(3)) to 7.7 $\mu$m PAH luminosity as a function of 24 $\mu$m continuum luminosity in molecular gas-rich galaxies. MOHEGs occupy the high-$L(\mathrm{H}_2)/L(\mathrm{PAH})$ regime above the threshold value of 0.04 (red triangle at lower limits), indicating dominant mechanical heating. The dashed sequence represents normal star-forming galaxies from the SINGS survey, which follow the expected correlation between PAH and H$_2$ emission for UV heating, adapted from \citep{Ogle2010}.}
    
    \label{fig:multigas-moheg}
\end{SCfigure}

Figure~2, illustrates how MOHEGs occupy a distinct region of parameter space characterised by enhanced $L(\mathrm{H}_2)/L(\mathrm{PAH})$ ratios  relative to normal star-forming galaxies with weak and compact radio jets. MOHEGs with large scale radio jets are clearly offset ($\ge$ 0.04 limit) from this sequence, indicating mechanical heating or turbulent dissipation rather than star formation. In contrast, normal star-forming galaxies (e.g. from the SINGS sample) show a tight correlation between H$_2$ and PAH emission consistent with UV heating. These properties indicate that mechanical energy injection from AGN jets and shocks dominates the excitation and dynamics of the molecular gas. The presence of LINERs, Seyferts, and dusty ellipticals within the same high-H$_2$/PAH regime further suggests that mechanical heating may be common among radio-loud AGN populations \citep{Kaneda2008,Ogle2010}. These results further support that mechanical energy from radio jets appears to drive turbulence and heat the molecular gas, leading to reduced star formation efficiency rather than complete quenching despite substantial molecular reservoirs. Thus multi-phase outflow diagnostics play an important role in probing AGN feedback and star formation regulation. For SKA~1 science, these systems offer a template for interpreting jet–ISM interactions, where combining radio continuum jet power, molecular line, and H~{\sc i} absorption data will enable population-wide tests of turbulence-dominated, low-efficiency star-forming reservoirs in radio-mode AGN feedback.

\subsection{Neutral Atomic Gas and Star Formation Suppression in Radio AGNs}
Neutral atomic hydrogen represents the intermediate phase in the baryon cycle connecting diffuse ionised gas and star-forming molecular reservoirs. It is a key tracer of atomic gas in the ISM and plays a crucial role in regulating the availability of cold gas for star formation. In radio-loud AGN, however, H~{\sc i} is primarily observed in absorption against compact radio continuum sources, making its observability strongly dependent on source morphology, orientation, and surface brightness. As a result, current samples are biased toward systems with compact cores or well-defined inner jets, limiting population-wide constraints on atomic gas content and kinematics in radio AGN. Molecular gas-rich galaxies such as MOHEGs, provide key reference systems in this context, in which  H~{\sc i} absorption, CO emission, jet structures and star forming properties can be directly compared. Furthermore due to their proximity to the dense cluster regions, the impact of environment can also be investigated for regulating star formation.
\begin{figure}[h]
    \centering
    \includegraphics[width=0.52\textwidth]{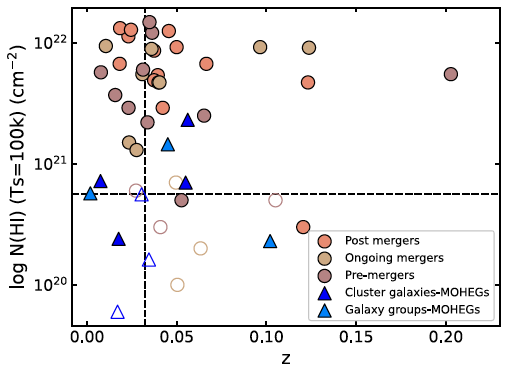}\hfill
    \includegraphics[width=0.45\textwidth]{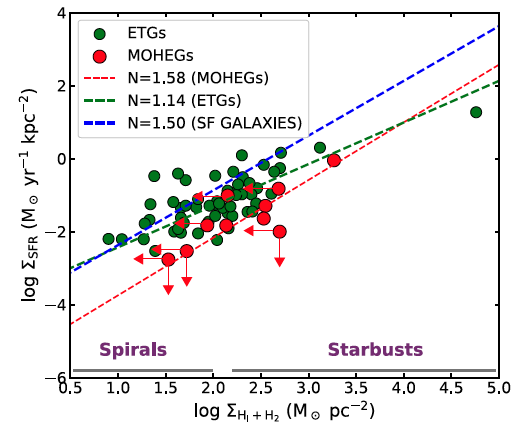}
\caption{
\textbf{Panel 1:} H~{\sc i} column density versus redshift for radio MOHEGs and merging galaxies from \citep{Dutta2019}, with unfilled markers representing non-detections. The dashed lines correspond to the median values extracted from the radio MOHEG sample, with a median column density of $N(\mathrm{H\,I}) \leq 5.65 \times 10^{20}\ \mathrm{cm^{-2}}$ at $z = 0.03245$ for a better representation of MOHEGs.
\textbf{Panel 2:} Kennicutt--Schmidt relation for normal star-forming galaxies, starbursts, early-type galaxies (ETGs), and molecular gas-rich galaxies. The gas surface density for molecular gas-rich systems is derived from H$_2$ rotational lines. These galaxies lie below the expected relation, indicating suppressed star formation efficiency \citep{Wagh2024}.
}
    \label{fig:HI_vs_mergers}
\end{figure}

Figure~3, compares the H~{\sc i} content and star formation efficiency of molecular gas-rich galaxies with other galaxy populations and environments. Panel~1 shows that these systems have systematically lower H~{\sc i} column densities than the merger sample of \citep{Dutta2019}, with a large fraction of non-detections. The median values, $z = 0.03245$ and $N(\mathrm{H\,I}) \leq 5.65 \times 10^{20}\ \mathrm{cm^{-2}}$, highlight a general deficiency of neutral atomic gas, suggesting that it may be depleted, ionised, or dynamically disrupted in dense environments. This supports a scenario in which mechanical feedback affects the survival and cooling of the atomic phase, thereby regulating the supply of gas available for molecular cloud formation. This behaviour supports a scenario in which mechanical feedback regulates the survival and cooling efficiency of the neutral phase, thereby controlling the supply of gas available for molecular cloud formation. 
Panel~2 shows the Kennicutt--Schmidt relation between gas surface density and star formation rate surface density. Star-forming galaxies and starbursts follow the canonical scaling relations \citep{Kennicutt2021}, while early-type galaxies (ETGs) follow the trends from \citep{Davis2014}. In contrast, molecular gas-rich galaxies lie systematically below these relations, indicating significantly reduced star formation efficiency at fixed gas surface density. Their molecular gas masses are derived from H$_2$ rotational line emission \citep{Ogle2010}, enabling a direct comparison with CO-based measurements in other populations. This offset implies that, despite retaining substantial molecular reservoirs, these galaxies form stars inefficiently.
A plausible explanation is that mechanical energy input from radio jets drives turbulence and heating across multiple gas phases, altering the density structure of the ISM and suppressing gravitational collapse. In this framework, the H~{\sc i}-H$_2$ transition is not governed solely by gas supply or environment, but is also strongly regulated by AGN-driven feedback.

The SKA will advance these studies by enabling sensitive, wide-band H~{\sc i} absorption surveys across large, statistically representative samples of radio-loud AGN. The SKA studies of radio AGN with large-scale jets will enable us to understand how cold gas cycles between atomic and molecular phases under the influence of mechanical feedback. In particular, whether jet--ISM interactions (i) heat and disrupt H~{\sc i}, suppressing its ability to cool into molecular gas, or (ii) compress atomic gas, enhancing the formation of H$_2$ and other molecular species such as CO, HCN, and HCO$^{+}$. This will enable direct measurements of atomic gas kinematics as a function of jet power, host galaxy properties, and evolutionary stage, extending current studies beyond small samples and probing the full chain of jet--ISM interaction in galaxies undergoing transitional phases, often characterised by shocked gas on multiple spatial scales (see Fig.~\ref{fig:gas-phase-moheg}). Further, H~{\sc i} absorption surveys with the SKA, when combined with CO, HCN, and HCO$^{+}$ observations, will provide a unified, multi-phase view of AGN-regulated star formation across cosmic time. Although this subsection focuses on low-redshift systems, the underlying physical processes are directly relevant to high-redshift radio galaxies. Elevated gas fractions, higher interstellar pressures, and more compact morphologies at earlier cosmic epochs may strengthen jet--ISM coupling, making studies of the H~{\sc i} to H$_2$ transition even more effective at higher redshifts.

\subsection{Jet-ISM Coupling and AGN Feedback}
In radio-loud AGNs, jets inject mechanical energy into the ISM, where it is dissipated via shocks, turbulence, and heating. The distribution of energy across atomic, molecular, and ionised phases determines how feedback regulates star formation. In this section, we examine how radio jets in rare class of molecular gas-rich FR I/II galaxies heat and disrupt the molecular gas and quench star formation. \citep{Ogle2010} showed that H$_2$ in MOHEGs likely originates from mergers or ICM cooling flows, and is heated by shocks and cosmic rays from AGN jets as discussed in previous sections. MOHEGs are used here primarily as nearby well-studied laboratories of jet–ISM interaction that motivate the broader SKA science case for statistical studies of cold gas feedback in radio-loud AGN populations. The presence of large-scale jets underscores the role of kinetic feedback, with radio power serving as a key tracer of the mechanical energy driving these effects.
\begin{figure}[ht]
    \centering
    \includegraphics[width=0.39\textwidth]{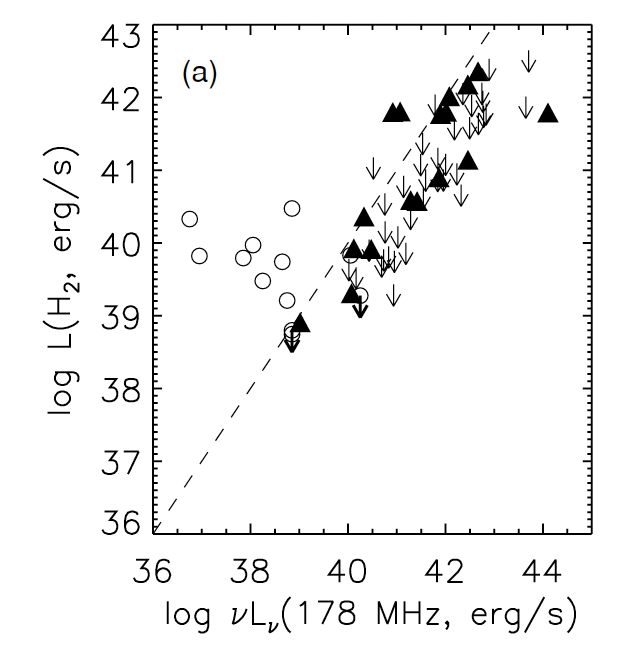} 
    \includegraphics[width=0.6\textwidth]{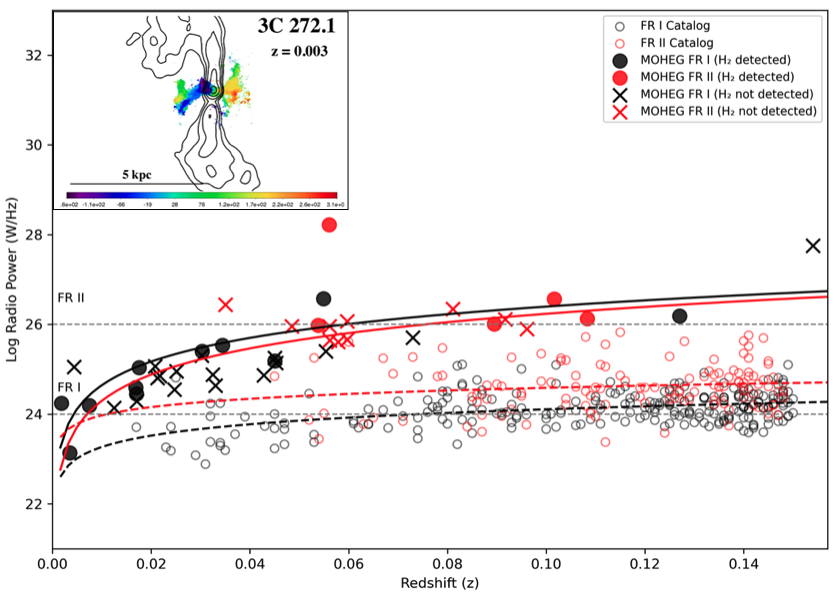} 
    \caption{\textbf{Panel 1:} H$_2$ luminosity (S(0)–S(3)) versus radio luminosity at 178 MHz. MOHEGs (triangles) lie linear to the star formation correlation line, indicating excess warm molecular hydrogen emission likely powered by mechanical heating \citep{Ogle2010}. Dusty ellipticals are shown in open circles \citep{Kaneda2008}.
    \textbf{Panel 2:} Inset-image: Kpc-scale radio jet seen in a typical molecular gas-rich galaxy 3C272.1 \citep{ Harrison2024}. In general, molecular gas-rich galaxies hosting FR I/FR II jets tend to show systematically enhanced 1.4 GHz radio power relative to the full galaxy population \citep{Capetti2017a,Capetti2017b}.}
    \label{fig:radio_feedback}
\end{figure}
Fig.~\ref{fig:radio_feedback} summarises the relationship between radio power and molecular hydrogen emission in molecular gas-rich galaxies. Panel~1 shows that molecular gas-rich galaxies exhibit significantly elevated H$_2$ luminosities, up to $10^{43}$ ergs/s, relative to their 178 MHz radio luminosities, far above the levels expected for star formation-driven systems (open circles). However, the correlation between warm H$_2$ luminosity and radio power remains weak and a large scatter in the ratio of $L(\mathrm{H}_2)/L_{178}$ is seen, likely reflecting variations in the molecular gas content, environmental density, and differences in the spatial and temporal scales traced by radio lobes versus H$_2$ emission. \citep{Ogle2010} further suggested that the compact (kpc-scale) H$_2$ regions may not overlap with the much more extended radio structures (up to hundreds of kpc), and the short cooling time measured for H$_2$ ($\sim$10$^4$ yr) gas can further complicate the link between instantaneous radio power and ongoing gas heating mechanisms within the ISM. This highlights the importance of time-dependent jet–ISM coupling in interpreting molecular emission diagnostics. Panel~2, shows that molecular gas-rich galaxies occupy the high end of the radio power distribution among FR I/II systems and exhibit indications of strong evolution in radio luminosity with redshift, reaching values $P_{1.4,\mathrm{GHz}} \gtrsim 10^{26}\,\mathrm{W\,Hz^{-1}}$. This suggests that molecular gas-rich galaxies trace a particularly active feedback phase in which jet kinetic energy deposition into the ISM is maximised. In contrast, non-molecular gas-rich FR II galaxies show a smoother evolution in radio power, consistent with less efficient coupling between jets and dense gas phases or perhaps gas-deficient radio AGN population \citep{Capetti2017a, Capetti2017b}. From a physical perspective, these differences likely reflect that in molecular gas-rich galaxies, dense gas-rich environments promote efficient conversion of kinetic jet power into turbulent and thermal energy within the molecular phase, enhancing heating and suppressing star formation. In gas-poor or lower-density systems, a larger fraction of jet energy may escape into the circumgalactic or intergalactic medium, reducing local ISM coupling efficiency (see Fig.~\ref{fig:gas-phase-moheg}). Further, these systems define an observational bridge between resolved local feedback and unresolved high-redshift galaxy populations. At higher redshifts, higher gas fractions and increased ambient pressures may enhance jet coupling efficiency, potentially shifting galaxies into regimes similar to or even more extreme than local molecular gas-rich galaxies.  Combined with CO and dense gas diagnostics from millimetre facilities, SKA radio continuum and absorption-line measurements will allow a unified description of how AGN jets regulate the thermal state, phase structure, and star formation efficiency of the ISM.

\subsection{Episodic AGN Feedback Activity and Cold Gas Regulation}
\noindent
Many radio AGNs show evidence of episodic jet activity, where multiple outbursts trace recurrent feedback cycles. These cycles regulate the ISM by repeatedly injecting energy, driving shocks and turbulence, and altering the H~{\sc i} - H$_2$ gas balance. A key diagnostic of episodic jet activity is spectral ageing of the radio continuum, which allows constraints on the timescales of active and remnant phases. Early studies of large-scale radio galaxies (e.g. \citep{Konar2006, Konar2012, Konar2013}) demonstrated that multiple jet episodes can be identified through spatial variations in spectral curvature and break frequencies. However, extending these techniques to compact and gas-rich systems has remained challenging due to sensitivity and resolution limitations. This regime is precisely where molecular gas-rich radio AGN such as NGC 1266, 3C 293, and IC 5063 become crucial, as they combine confined radio structures with rich atomic and molecular gas reservoirs that directly respond to jet energy injection. The SKA will enable population-wide spectral ageing studies across large populations of radio AGN, allowing the separation of active, relic, and restarted jet components even in low-surface-brightness emission \citep{Hardcastle01.2026.SKA}. In combination with multi-phase gas tracers (CO, OH, and H~{\sc i} absorption), this will provide a spatially resolved connection between jet history and cold gas regulation. Complementary molecular gas observations from facilities such as ALMA will further constrain the distribution and excitation of the H$_2$-traced gas, enabling a full multi-phase view of the ISM response to AGN duty cycles. This framework directly connects jet-driven feedback in local molecular gas-rich galaxies with molecular gas evolution in high-redshift galaxies, where gas fractions are higher and turbulence-driven star formation regulation dominates \citep{Tacconi2020}.

\section{SKA Prospects for Statistical Studies of AGN Feedback  and multi-wavelength synergies}
\noindent
A comprehensive understanding of molecular-gas rich AGN feedback requires a statistically representative sample of radio AGN spanning a wide range of environments, gas contents, and redshifts. In previous sections, we have shown that molecular gas-rich radio galaxies provide direct evidence that jet-driven mechanical feedback strongly modifies the thermal state, turbulence, and star formation efficiency of the multi-phase ISM. However, current observational constraints remain limited by sensitivity, spatial resolution, and incomplete spectral coverage across atomic and molecular gas phases. The SKA will enhance this work by enabling sensitive, high-resolution studies of both radio continuum and spectral line emission over a broad frequency range (50 MHz–15 GHz for SKA~1).  Its combination of $\mu$Jy sensitivity, sub-arcsecond imaging, and wide instantaneous bandwidth will allow simultaneous constraints on jet energetics, spectral ageing, and cold gas properties, providing complementary constraints through absorption and selected low-J transitions in a bright sample of AGNs. In particular, the SKA will reach radio powers down to $P_{1.5\,\mathrm{GHz}} \sim 10^{18}$ W Hz$^{-1}$, enabling the detection of faint, extended jet structures and low-surface-brightness relic emission associated with past feedback cycles.

\begin{table}[ht]
\centering
\small
\begin{tabular}{lccccp{3.0cm}}
\hline
Transition & Rest Freq. & Obs. Freq. & SKA1 Band & Redshift Range & Science Goal \\
 & (GHz) & (GHz) &  &  &  \\
\hline

H~{\sc i} 21-cm 
& 1.420 
& 1.42 -- 0.35 
& Band 1/2 (0.35--1.76 GHz) 
& $0 \lesssim z \lesssim 3$ 
& Gas inflows/outflows\\

OH 18-cm 
& 1.612--1.667 
& 1.67 -- 0.40 
& Band 1/2 (0.35--1.76 GHz) 
& $0 \lesssim z \lesssim 3$ 
& Maser, shocked gas \\

CH$_3$OH
& 6.668 
& 6.7 -- 4.6 
& Band 5 (4.6--15.3 GHz) 
& $0 \lesssim z \lesssim 0.45$ 
& Maser, shocked gas\\

H$_2$CO 
& 4.829 
& 4.8 -- 4.6 
& Band 5 (4.6--15.3 GHz) 
& $z \lesssim 0.05$ 
& Dense molecular gas\\

RRLs
& $\sim$1--10 
& $\sim$10 -- 1 
& Band 2--5 
& $0 \lesssim z \lesssim 2$ 
& Ionised gas \\

CO(1--0)$^{a}$ 
& 115.271 
& $\lesssim 15$ 
& Upper Band 5$^{b}$ 
& $z \gtrsim 6.7$ 
& Cold molecular gas$^{c}$\\

HCN(1--0)$^{a}$ 
& 88.632 
& $\lesssim 15$ 
& Upper Band 5$^{b}$  
& $z \gtrsim 4.9$ 
& Dense molecular gas \\

HCO$^{+}$(1--0)$^{a}$ 
& 89.188 
& $\lesssim 15$ 
& Upper Band 5$^{b}$ 
& $z \gtrsim 5.0$ 
& Shocked dense gas \\
\hline
\end{tabular}
\caption{
Key spectral tracers relevant for AGN feedback studies accessible with SKA~1 and future high-frequency extensions. The table lists rest-frame frequencies, approximate observable frequency ranges 
($\nu_{\rm obs}=\nu_{\rm rest}/(1+z)$), corresponding SKA1-MID bands, indicative redshift coverage, and primary science applications. 
H~{\sc i} and OH transitions will be accessible across a broad redshift range in both emission and absorption. 
$^a$: Higher-frequency molecular transitions such as CO(1--0), HCN(1--0), and HCO$^{+}$(1--0) become observable only at high redshift when redshifted into the highest SKA frequency bands and are expected to be detectable primarily in absorption against bright radio continuum sources or in rare extreme systems.
$^{b}$: Future high-frequency extensions. $^{c}$ Rare high-$z$ bright systems.
The quoted redshift limits are indicative and depend on the final SKA frequency configuration and survey sensitivity.
}
\label{tab:ska_lines_table}
\end{table}
\begin{figure}[h]
    \centering
    \includegraphics[width=0.9\textwidth]{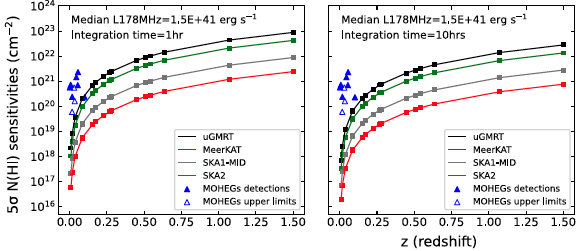}
    \caption{\textbf{Panel 1 and 2:} N(H~{\sc i}) sensitivities for SKA and pathfinders w.r.t to redshift to detect H~{\sc i} in molecular gas-rich galaxies for different integration times \citep{Wagh2024}. The molecular gas-rich galaxies sample discussed in previous sections is represented by blue triangles.
    }
    \label{fig:ska_HI}
\end{figure}

As illustrated in Table~1 and Fig.~5, SKA~1 will extend direct measurements of atomic gas inflows and outflows in AGN host galaxies out to $z \sim 3$, well beyond currently studied samples limited to $z \sim 0.2$. This will enable statistically robust samples at lower redshifts and provide key diagnostics of jet--ISM interactions across a much broader redshift range. Broad, asymmetric, and blueshifted absorption profiles will trace jet-driven acceleration of neutral gas, while narrow components will probe more quiescent reservoirs. Furthermore, the SKA~1 will enable observations of neutral atomic hydrogen via the 21-cm line absorption studies extending up to $z \sim 6$, subject to the presence of sufficiently bright background radio continuum sources. With sensitivity reaching column densities as low as $\sim10^{18}$ cm$^{-2}$ and sub-kiloparsec resolution at moderate redshifts ($z \lesssim 0.5$), the SKA~1 will open up detailed H~{\sc i} absorption studies in molecular gas-rich galaxies, as illustrated in Fig. 5, \citep{Wagh2024}. 
Only selected low-J molecular transitions become accessible within SKA frequency bands at high redshift, making these observations complementary to facilities such as ALMA and ngVLA. These capabilities are crucial for tracing neutral gas inflows-outflows, a key diagnostic of AGN-driven feedback \citep{Morganti2018}. Additionally, the SKA’s wide instantaneous bandwidth, VLBI-scale resolution of $\sim$subarcsec at 1.4 GHz, and high dynamic range will allow the detection of kpc-scale H~{\sc i} clouds and broad line profiles, capturing the high-velocity wings associated with jet- or wind-driven outflows \citep{Pandey-Pommier03.2026.SKA}. Stacking techniques will further enhance sensitivity to faint, diffuse H~{\sc i} features in the ISM, such as cold gas clouds and filamentary inflows, expanding our understanding of gas cycling in active galaxies.

Beyond neutral hydrogen, the SKA~1’s broad frequency coverage will allow access to a range of additional spectral lines critical for probing dense molecular regions within the ISM, as listed in Table~1. These include CO (at very high redshift), OH, HCN, $HCO^{+}$, and $CH_{3}OH$, which trace dense molecular gas, dense shocked regions, and maser activity, respectively \citep{saintonge2022}. 

With its sub-arcsecond resolution, the SKA will achieve parsec-scale resolution at low redshifts ($z\le0.5$) and potentially resolve sub-kpc structures in bright high-redshift systems in its highest frequency bands (Band 5a/b), critical for investigating the interaction between radio jets and molecular gas in regions such as shocks, dense clouds, or disrupted filaments. While many of these lines have been inaccessible or marginally detectable with current facilities due to sensitivity and frequency limitations, the SKA~1’s broad spectral coverage ($\ge$ 15 GHz) and $\mu$Jy-level sensitivity will improve the detectability of selected molecular transitions primarily in absorption and in rare bright systems, \citep{vanLoon01.2026.SKA}. Further, as the table reveals, line detectability at high redshift is limited due to the sensitivity and spatial resolution required to isolate cloud structures. For instance, in SKA Band 1 and 2, H~{\sc i}, and $CH_{3}OH$ at $z>2$ may become accessible in favourable systems; however, OH lines at $z\le2$ fall within the achievable resolution limits of the detection of sub-kiloparsec clouds. In contrast, at high frequency bands (Band 5a/b), the SKA~1 will resolve structures on scales of a few hundred parsecs at moderate redshifts and sub-kpc scales at high redshift in its highest-frequency bands making these bands optimal for targeting high-redshift CO and $HCO^{+}$ transitions arising from dense molecular clouds and jet-driven shocked regions. These observations are crucial for characterizing feedback-regulated star formation and the abundance of dense gas regions at higher redshifts. Interpretation of multi-phase gas diagnostics nevertheless remains subject to several uncertainties, including spin-temperature assumptions in H~{\sc i} absorption measurements, CO-to-$H_{2}$ conversion factor variations in shocked environments, and selection biases toward radio-bright AGN.

Finally, multi-wavelength and synergetic observations with the SKA~1, will greatly enhance such studies by probing the feedback processes in molecular gas-rich galaxies, positioning them as key laboratories for understanding the full cycle of galaxy evolution, from star formation to AGN-driven activity. Facilities such as ALMA and NOEMA provide high-resolution imaging of dense molecular gas (such as CO, HCN), complementing SKA’s (H~{\sc i}, OH) line capabilities. Together, these enable a complete view of the cold ISM across molecular and atomic phases. Infrared missions like \textit{JWST} will reveal warm molecular hydrogen and dust emission, essential for probing the shocked ISM and obscured star formation in molecular gas-rich galaxies \citep{saintonge2022}. Optical and X-ray telescopes (e.g., VLT/MUSE, \textit{Chandra}, \textit{Athena}) provide measurements of ionized gas kinematics and AGN energetics, linking feedback signatures to the central engine. 

\section*{Summary}
In this chapter, we have evaluated the potential of SKA~1 to investigate AGN-driven feedback in radio-loud galaxies hosting rich multi-phase interstellar media, with particular emphasis on molecular gas-rich systems as nearby benchmarks of jet--ISM interaction. Rather than proposing a dedicated survey exclusively targeting MOHEGs, we use these well-characterized local systems as physically motivated laboratories for understanding the broader impact of radio jets on cold gas reservoirs. Their observed combination of disturbed molecular gas, H~{\sc i} absorption, and suppressed star formation provides an empirical framework for interpreting future SKA studies of statistically significant samples of radio-loud AGN spanning a wide range of environments, radio powers, and evolutionary stages. Our analysis shows that the SKA will enable detailed studies of neutral atomic hydrogen via H~{\sc i} absorption and emission, tracing gas inflows, outflows, and disturbed kinematics from the local Universe out to intermediate redshifts of $z \sim 2$ in emission and up to $z \sim 6$ in absorption, thereby providing key constraints on the atomic gas reservoir in AGN host galaxies across cosmic time. In parallel, SKA continuum observations will map extended and low-surface-brightness radio jets over a broad redshift range, allowing direct reconstruction of jet energetics, spectral ageing, and their coupling to the surrounding interstellar medium.

At higher frequencies, access to low-J molecular transitions such as CO(1–0), HCN(1–0), and HCO$^{+}$(1–0) will enable SKA~1 to probe the cold molecular gas content and dense gas fractions in rare systems in absorption against bright radio continuum AGN host galaxies from the local Universe up to $z \sim 5$ for dense gas tracers and $z \sim 7$ for CO(1–0) in the most extreme star-forming systems. This will link jet-driven feedback to the suppression, enhancement, or redistribution of star-forming gas across cosmic time. These capabilities will allow us to extend molecular gas-rich galaxy-based physical calibrations of jet–ISM coupling to larger and more diverse galaxy populations, bridging the gap between local multi-phase feedback studies and high-redshift galaxy evolution.

Overall, SKA~1 observations will provide a unified framework connecting radio jets, atomic gas, and molecular gas, enabling a physically consistent description of how AGN regulate the cold interstellar medium and star formation efficiency from the nearby Universe to high redshift. In synergy with ALMA, NOEMA, JWST, and future optical and X-ray facilities such as the ELT, SKA will play a central role in advancing a multi-wavelength, multi-phase understanding of galaxy evolution driven by mechanical AGN feedback. The high-frequency SKA observations will be complementary to the ngVLA, for molecular line studies and large-area surveys (e.g., SPT, ACT, and WST), enabling statistically robust studies of AGN-driven cold gas regulation across cosmic time.

\section*{Acknowledgment}
We thank the referee for reviewing this manuscript and for providing comments during the evaluation process. We are especially grateful to the members of the Editorial Board for their prompt handling of the manuscript and their timely support, which facilitated its timely publication.
\section*{Author Ordering}
Authors for this chapter are ordered according to their overall level of contribution, in line with that expected for a small author list publication.

\bibliographystyle{abbrvnat-maxbibnames4}
\bibliography{chapter} 

@incollection{RanWang01.2026.SKA, author = {Ran Wang and author2 and author3 and author4 and author5},title = {Cm-wavelength Studies of Molecular Gas and Star Formation at High Redshift with the SKA },year = {2026},publisher = {Advancing Astrophysics with the SKA},note = {arXiv search: Report number AASKAII/RanWang01},booktitle = {Advancing Astrophysics with the SKA -- II (AASKAII)}}

@incollection{vanLoon01.2026.SKA, author = {Jacco van Loon and author2 and author3 and author4 and author5},title = {SKA Extragalactic spectral lines review chapter},year = {2026},publisher = {Advancing Astrophysics with the SKA},note = {arXiv search: Report number AASKAII/vanLoon01},booktitle = {Advancing Astrophysics with the SKA -- II (AASKAII)}}

@incollection{Pandey-Pommier03.2026.SKA, author = {Mamta Pandey-Pommier and author2 and author3 and author4 and author5},title = {Uncovering neutral Hydrogen clouds in Radio Galaxies in the SKA era},year = {2026},publisher = {Advancing Astrophysics with the SKA},note = {arXiv search: Report number AASKAII/Pandey-Pommier03},booktitle = {Advancing Astrophysics with the SKA -- II (AASKAII)}}

@incollection{Hardcastle01.2026.SKA, author = {Martin J. Hardcastle and author2 and author3 and author4 and author5},title = {Radio Galaxies and Jet Duty Cycles},year = {2026},publisher = {Advancing Astrophysics with the SKA},note = {arXiv search: Report number AASKAII/Hardcastle01},booktitle = {Advancing Astrophysics with the SKA -- II (AASKAII)}}

@article{Capetti2017a,
  author  = {Capetti, A. and Massaro, F. and Baldi, R. D.},
  title   = {FRICAT: A FIRST catalog of FR I radio galaxies},
  journal = {Astronomy \& Astrophysics},
  volume  = {598},
  pages   = {A49},
  year    = {2017},
  doi     = {10.1051/0004-6361/201629287}
}

@article{Capetti2017b,
  author  = {Capetti, A. and Massaro, F. and Baldi, R. D.},
  title   = {FRIICAT: A FIRST catalog of FR II radio galaxies},
  journal = {Astronomy \& Astrophysics},
  volume  = {601},
  pages   = {A81},
  year    = {2017},
  doi     = {10.1051/0004-6361/201630247}
}

@ARTICLE{Konar2006,
       author = {{Konar}, C. and {Saikia}, D. and et. al.},
        title = "{Spectral ageing analysis of the double-double radio galaxy J1453+3308}",
      journal = {MNRAS},
         year = 2006,
        month = Oct,
       volume = {372},
        pages = {693-702},
          doi = {10.1111/j.1365-2966.2006.10874.x},
       adsurl = {https://ui.adsabs.harvard.edu/abs/2006MNRAS.372..693K/abstract}
}

@ARTICLE{Konar2012,
       author = {{Konar}, C. and {Hardcastle}, M. and et. al.},
        title = "{Rejuvenated radio galaxies J0041+3224 and J1835+6204: how long can the quiescent phase of nuclear activity last?}",
      journal = {MNRAS},
         year = 2012,
        month = Aug,
       volume = {424},
        pages = {1061 – 1076},
          doi = {10.1111/j.1365-2966.2012.21279.x},
       adsurl = {https://ui.adsabs.harvard.edu/abs/2012MNRAS.424.1061K/abstract}
}

@ARTICLE{Konar2013,
       author = {{Konar}, C. and {Hardcastle}, M. and et. al.},
        title = "{Episodic radio galaxies J0116-4722 and J1158+2621: can we constrain the quiescent phase of nuclear activity? }",
      journal = {MNRAS},
         year = 2013,
        month = Apr,
       volume = {430},
        pages = {2137 – 2153},
          doi = {10.1093/mnras/stt040},
       adsurl = {https://ui.adsabs.harvard.edu/abs/2013MNRAS.430.2137K/abstract }
}

@ARTICLE{boselli2006,
       author = {{Boselli}, A. and {Gavazzi}, G.},
        title = "{Environmental Effects on Late-Type Galaxies in Nearby Clusters}",
      journal = { Publications of the Astronomical Society of the Pacific},
         year = 2006,
        month = apr,
       volume = {118},
        pages = {517-559},
          doi = {10.1086/500691},
       adsurl = {https://ui.adsabs.harvard.edu/abs/2006PASP..118..517B/abstract}
}

@ARTICLE{kennicutt2012,
       author = {{Kennicutt}, R.~C. and {Evans}, N.~J.},
        title = "{Star Formation in the Milky Way and Nearby Galaxies}",
      journal = {\araa},
         year = 2012,
        month = sep,
       volume = {50},
        pages = {531-608},
          doi = {10.1146/annurev-astro-081811-125610},
       adsurl = {https://ui.adsabs.harvard.edu/abs/2012ARA&A..50..531K}
}

@ARTICLE{saintonge2022,
       author = {{Saintonge}, A. and {Catinella}, B.},
        title = "{Cold Gas in Galaxies: The Key Ingredient Driving Galaxy Evolution}",
      journal = {\araa},
         year = 2022,
        month = Aug,
       volume = {60},
        pages = {319-364},
          doi = {10.1146/annurev-astro-021022-043545},
       adsurl = {https://ui.adsabs.harvard.edu/abs/2022ARA&A..60..319S}
}

@ARTICLE{krumholz2013,
       author = {{Krumholz}, M.~R.},
        title = "{The big problems in star formation: The star formation rate, stellar clustering, and the initial mass function}",
      journal = {Physics Reports},
         year = 2014,
        month = Jun,
       volume = {539},
        pages = {49-134},
          doi = {10.48550/arXiv.1402.0867},
       adsurl = {https://arxiv.org/abs/1402.0867}
}

@ARTICLE{leroy2008,
       author = {{Leroy}, A.~K. and {Walter}, F. and {Brinks}, E. and et al.},
        title = "{The Star Formation Efficiency in Nearby Galaxies: Measuring Where Gas Forms Stars Effectively}",
      journal = {The Astronomical Journal},
         year = 2008,
        month = dec,
       volume = {136},
        pages = {2782-2845},
          doi = {10.1088/0004-6256/136/6/2782},
       adsurl = {https://ui.adsabs.harvard.edu/abs/2008AJ....136.2782L}
}

@ARTICLE{fernandez2016,
       author = {{Fern{\'a}ndez}, X. and {Hansung}, G.-. and {van Gorkom}, J. and et al.},
        title = "{HIGHEST REDSHIFT IMAGE OF NEUTRAL HYDROGEN IN EMISSION: A CHILES DETECTION OF A STARBURSTING GALAXY AT z = 0.376}",
      journal = {The Astrophysical Journal Letters},
         year = 2016,
        month = apr,
       volume = {824},
       number = {1},
        pages = {L1},
          doi = {10.3847/2041-8205/824/1/L1},
       adsurl = {https://ui.adsabs.harvard.edu/abs/2016ApJ...824L...1F/abstract}
}

@INPROCEEDINGS{Morganti2015,
       author = {{Morganti}, R. and {Sadler}, E.~M. and {Curran}, S.~J.},
        title = "{HI absorption in galaxies and AGN: probing the inner regions with the SKA}",
     booktitle = {Advancing Astrophysics with the Square Kilometre Array (AASKA14)},
         year = 2015,
        month = apr,
          eid = {134},
       adsurl = {https://ui.adsabs.harvard.edu/abs/2015aska.confE.134M}
}

@ARTICLE{aditya2018,
       author = {{Aditya}, J.~N.~H.~S. and {Kanekar}, N.},
        title = "{HI 21 cm absorption at z = 1.275: The highest redshift detection of associated HI 21 cm absorption}",
      journal = {MNRAS},
         year = 2018,
        month = Jan,
       volume = {473},
        pages = {59-67},
          doi = {10.1093/mnras/stx2325},
       adsurl = {https://ui.adsabs.harvard.edu/abs/2018MNRAS.473...59A/abstract}
}

@article{Volk1996,
  author    = {Völk, H. and Aharonian, F. and et al.},
  title     = {The Nonthermal Energy Content and Gamma Ray Emission of Starburst Galaxies and Clusters of Galaxies},
  journal = {Space Science Reviews},
  pages     = {279--297},
  year      = {1996},
  volume    = {75},
  doi       = {10.1007/BF00195040},
  adsurl    = {https://adsabs.harvard.edu/full/1996SSRv...75..279V},
}

@ARTICLE{cicone2014,
       author = {{Cicone}, C. and {Maiolino}, R. and {Sturm}, E. and et al.},
        title = "{Massive molecular outflows and evidence for AGN feedback from CO observations}",
      journal = {\aap},
         year = 2014,
        month = feb,
       volume = {562},
        pages = {A21},
          doi = {10.1051/0004-6361/201322464},
       adsurl = {https://ui.adsabs.harvard.edu/abs/2014A&A...562A..21C}
}

@ARTICLE{morganti2018,
       author = {{Morganti}, R. and {Oosterloo}, T.},
        title = "{The interstellar and circumnuclear medium of active nuclei
traced by HI 21-cm absorption}",
      journal = {\aapr},
         year = 2018,
        month = Jul,
        pages = {60},
          doi = {10.1007/s00159-018-0109-x},
       adsurl = {https://ui.adsabs.harvard.edu/abs/2018A%26ARv..26....4M/abstract}
}

@ARTICLE{Castignani2020,
       author = {{Castignani}, G. and {Pandey-Pommier}, M. and {Hamer}, S.~L. and et al.},
        title = "{Molecular gas in CLASH brightest cluster galaxies at z ~ 0.2–0.9}",
      journal = {\aap},
         year = 2020,
       volume = {640},
        pages = {A65},
          doi = {10.1051/0004-6361/202038081},
archivePrefix = {arXiv},
       eprint = {2006.00019},
 primaryClass = {astro-ph.GA},
       adsurl = {https://ui.adsabs.harvard.edu/abs/2020A&A...640A..65C}
}

@ARTICLE{oosterloo2017,
       author = {{Oosterloo}, T. and Oonk, R. and {Morganti}, R. and et al.},
        title = "{Properties of the molecular gas in the fast outflow in the Seyfert galaxy IC 5063}",
      journal = {Astronomy \& Astrophysics},
         year = 2017,
        month = dec,
       volume = {608},
        pages = {A38},
          doi = {10.1051/0004-6361/201731781},
       adsurl = {https://ui.adsabs.harvard.edu/abs/2017A&A...608A..38O}
}

@ARTICLE{hardcastle2007,
       author = {{Hardcastle}, M.~J. and {Evans}, D.~A. and {Croston}, J.~H.},
        title = "{ Hot and cold gas accretion and feedback in radio-loud active galaxies }",
      journal = {\mnras},
         year = 2007,
        month = apr,
       volume = {376},
        pages = {1849-1856},
          doi = {doi:10.1111/j.1365-2966.2007.11572.x},
       adsurl = {https://ui.adsabs.harvard.edu/abs/2007MNRAS.376.1849H}
}

@ARTICLE{Carilli2013,
       author = {{Carilli}, C.~L. and {Walter}, F.},
        title = "{Cool Gas in High-Redshift Galaxies}",
      journal = {\araa},
         year = 2013,
        month = sep,
       volume = {51},
        pages = {105-161},
          doi = {10.1146/annurev-astro-082812-140953},
       adsurl = {https://ui.adsabs.harvard.edu/abs/2013ARA&A..51..105C}
}

@ARTICLE{Tacconi2020,
       author = {{Tacconi}, L.~J. and {Genzel}, R. and {Sternberg}, A.},
        title = "{The Evolution of the Star-forming Interstellar Medium across Cosmic Time}",
      journal = {\araa},
         year = 2020,
        month = aug,
       volume = {58},
        pages = {157-203},
          doi = {10.1146/annurev-astro-082812-141034},
       adsurl = {https://ui.adsabs.harvard.edu/abs/2020ARA&A..58..157T}
}

@ARTICLE{harrison2024,
  author = {{Harrison}, C.~M. and {Ramos Almeida}, C.},
  title = "{Observational Tests of Active Galactic Nuclei Feedback: An Overview of Approaches and Interpretation}",
  journal = {Galaxies},
  year = {2024},
  volume = {12},
  number = {2},
  pages = {17},
  doi = {10.3390/galaxies12020017},
  url = {https://doi.org/10.3390/galaxies12020017}
}

@ARTICLE{Darling2001,
       author = {{Darling}, J. and Riccardo, G.},
        title = "{A Search for OH Megamasers at Z>0.1. II. Further Results}",
      journal = {\apj},
         year = 2001,
        month = Mar,
       volume = {121},
        pages = {1278-1293},
          doi = {10.1086/319413},
       adsurl = {arXiv:astro-ph/0102345}
}

@ARTICLE{Darling2007,
       author = {{Darling}, J.},
        title = "{A Dense Gas Trigger for OH Megamasers}",
      journal = {\apj},
         year = 2007,
        month = Oct,
       volume = {669},
        pages = {L9-L12},
          doi = {10.1086/523756},
       adsurl = {https://ui.adsabs.harvard.edu/abs/2007ApJ...669L...9D/abstract}
}

@ARTICLE{Bertoldi2003,
       author = {Bertoldi, F. and Cox, P. and Neri, R.},
        title = "{High-excitation CO in a quasar host galaxy at z=6.42}",
      journal = {Astronomy and Astrophysics},
         year = 2003,
        month = Sep,
       volume = {409},
        pages = {L47–L50},
          doi = {10.1051/0004-6361:20031345 },
       adsurl = {https://ui.adsabs.harvard.edu/abs/2003A%26A...409L..47B/abstract}
}

@ARTICLE{Riechers2007,
       author = {{Riechers}, D.~A. and {Walter}, F. and {Carilli}, C.~L. and et al.},
        title = "{OBSERVATIONS OF DENSE MOLECULAR GAS IN A QUASAR HOST GALAXY AT z p 6.42: FURTHER EVIDENCE FOR A NONLINEAR DENSE GAS–STAR FORMATION RELATION AT EARLY COSMIC TIME}",
      journal = {\apj},
         year = 2006,
        month = dec,
       volume = {671},
        pages = {L13–L16},
          doi = {10.1086/524871},
       adsurl = {https://ui.adsabs.harvard.edu/abs/2007ApJ...671L..13R/abstract}
}

@ARTICLE{Marshall2017,
       author = {Marshall, M. and Ellingsen, S. and Lovell, J. and et al.},
        title = "{Methanol absorption in PKS B1830-211 at milliarcsecond scales }",
      journal = {MNRAS},
         year = 2017,
        month = Apr,
       volume = {466},
        pages = {2450-2457},
          doi = {10.1093/mnras/stw3295},
       adsurl = {https://ui.adsabs.harvard.edu/abs/2017MNRAS.466.2450M/abstract}
}

@ARTICLE{Impellizzeri2008,
       author = {{Impellizzeri}, C. and {Henkel},C. and {Roy}, A. and et al.},
        title = "{6.7 GHz methanol absorption toward the Seyfert 2 galaxy NGC 3079}",
      journal = {Astronomy and Astrophysics},
         year = 2008,
        month = sep,
       volume = {484},
        pages = {L43-L46},
          doi = {10.1051/0004-6361:200809985},
       adsurl = {https://arxiv.org/abs/0805.1063}
}

@article{Ogle2010,
  author = {Ogle, P. M. and Boulanger, F. and Guillard, P. and Evans, D. A. and Antonucci, R. and Appleton, P. N. and Nesvadba, N. and Leipski, C.},
  title        = {Powerful H$_2$ Line Cooling in Stephan's Quintet II: Warm Molecular Hydrogen in Galaxies},
  journal      = {Astrophysical Journal},
  year         = {2010},
        month = Nov,
       volume = {724},
       number = {2},
        pages = {1193-1209},
          doi = {10.1088/0004-637X/724/2/1193},
  archivePrefix= {arXiv},
  primaryClass = {astro-ph.GA},
  adsurl       = {https://arxiv.org/abs/1009.4533}
}

@article{veilleux2020,
  author    = {Veilleux, S. and Maiolino, R. and Bolatto, A. D. and Aalto, S.},
  title     = {Cool outflows in galaxies and their implications},
  journal   = {The Astronomy and Astrophysics Review},
  volume    = {28},
  number    = {1},
  pages     = {2},
  year      = {2020},
  doi       = {10.1007/s00159-019-0121-9}
}

@article{Kaneda2008,
  author    = {Kaneda, H. and Onaka, T. and Sakon, I. and et al.},
  title     = {Properties of Polycyclic Aromatic Hydrocarbons in Local Elliptical Galaxies Revealed by the Infrared Spectrograph on Spitzer},
  journal   = {The Astrophysical Journal},
  year      = {2008},
  volume    = {684},
  number    = {1},
  pages     = {270-278},
  doi       = {10.1086/590243}
}

@article{Wagh2024,
  author       = {Wagh, S. and Pandey-Pommier, M and Roy, Nirupam and others},
  title        = {Exploring neutral hydrogen in the radio Molecular Hydrogen Emission Galaxies (MOHEGs) and prospects with the SKA},
  journal      = {The Astrophysical Journal},
  year         = {2024},
        month = Mar,
       volume = {963},
       number = {2},
        pages = {11pp},
          doi = {10.3847/1538-4357/ad1edf},
archivePrefix = {arXiv},
  eprint       = {2401.07613},
  archivePrefix= {arXiv},
  primaryClass = {astro-ph.GA},
  adsurl       = {https://arxiv.org/abs/2401.07613}
}

@article{Dutta2019,
  author    = {Dutta, R. and Srianand, R. and Gupta, N.},
  title     = {Prevalence of neutral gas in centres of merging galaxies−II: nuclear H I
and multiwavelength properties},
  journal   = {Monthly Notices of the Royal Astronomical Society},
  volume    = {489},
  pages     = {1099–1109},
  year      = {2019},
  doi       = {10.1093/mnras/stz2178}
}

@article{Kennicutt2021,
  author    = {Kennicutt, Robert C. and De Los Reyes, M. A. C.},
  title     = {Star Formation Laws in Galaxies: A Revision and Extension of the Schmidt Law},
  journal   = {The Astrophysical Journal},
  year      = {2021},
  volume    = {908},
  number    = {1},
  pages     = {61},
  doi       = {10.3847/1538-4357/abd3a2}
}

@article{Davis2014,
  author    = {Davis, T. A. and Young, L. M. and Crocker, A. F. and et al.},
  title     = {The ATLAS$^{\mathrm{3D}}$ Project – XX. Mass–size and mass–σ distributions of early-type galaxies: bulge fraction drives kinematics, mass–size and mass–σ relations},
  journal   = {Monthly Notices of the Royal Astronomical Society},
  year      = {2014},
  volume    = {444},
  number    = {4},
  pages     = {3427--3446},
  doi       = {10.1093/mnras/stu570}
}

\end{document}